\documentclass[aps,prd,superscriptaddress,showpacs,nofootinbib,floatfix]{revtex4}
\usepackage{epsfig,dcolumn,bm}

\usepackage{graphicx}
\usepackage{amsmath}
\usepackage{rotating}

\def\tstrut{\vrule height2.5ex depth0pt width0pt} 

\newcommand{\Db}{{\bar{D}}}

\newcommand{\Dsb}{{\bar{D}_s}}

\newcommand{\DSb}{{\bar{D}^*}}

\newcommand{\DsSb}{{\bar{D}_s^*}}

\newcommand{\Nucleon}{{N}}

\newcommand{\SigmaS}{{\Sigma^*}}

\newcommand{\Cascada}{{\Xi}}
\newcommand{\CascadaS}{{\Xi^*}}

\newcommand{\lc}{{\cal L}}

\newcommand{\oh}{{1\over2}}
\newcommand{\trh}{{3\over2}}

\newcommand{\be}{\begin{eqnarray}}
\newcommand{\ee}{\end{eqnarray}}
\newcommand{\nn}{\nonumber}

\newcommand{\Eq}[1]{Eq.~(\ref{#1})}

\newcommand{\del}{\Delta}
\newcommand{\sigs}{{\Sigma^*}}
\newcommand{\css}{\Xi^*}
\newcommand{\db}{\bar{D}}
\newcommand{\dsb}{\bar{D}_s}
\newcommand{\dv}{D^*}
\newcommand{\dvb}{\bar{D}^*}

\newcommand{\dsvb}{\bar{D}_s^*}

\begin{document}

\title{Exotic dynamically generated baryons
 with negative charm quantum number}

\author{D.~Gamermann}{\thanks{E-mail: daniel.gamermann@ific.uv.es}}
\affiliation{Instituto de F\'isica corpuscular (IFIC), Centro Mixto
Universidad de Valencia-CSIC,\\ Institutos de Investigaci\'on de
Paterna, Aptdo. 22085, 46071, Valencia, Spain}
\affiliation{Departament d'Estructura i Constituents de la 
Mat\`eria and Institut de Ci\`encies del Cosmos,
Universitat de Barcelona, Avda. Diagonal 647, E-08028 Barcelona, Spain}

\author{C.~Garc\'ia-Recio}{}
\affiliation{Departamento de F\'isica At\'omica, Molecular y 
Nuclear, Universidad de Granada, E-18071 Granada, Spain}

\author{J.~Nieves}{\thanks{E-mail: jmnieves@ific.uv.es}}
\affiliation{Instituto de F\'isica corpuscular (IFIC), Centro Mixto
Universidad de Valencia-CSIC,\\ Institutos de Investigaci\'on de
Paterna, Aptdo. 22085, 46071, Valencia, Spain}

\author{L.L.~Salcedo}{}
\affiliation{Departamento de F\'isica At\'omica, 
Molecular y Nuclear, Universidad de Granada, E-18071 Granada, Spain}

\author{L.~Tolos}
\affiliation{Theory Group, KVI, University of Groningen, 
Zernikelaan 25, 9747 AA Groningen, The Netherlands}

\begin{abstract}
Following a model based on the $\rm{SU}(8)$ symmetry that treats heavy
pseudo-scalars and heavy vector mesons on an equal footing, as
required by heavy quark symmetry, we study the interaction of baryons
and mesons in coupled channels within an unitary approach that
generates dynamically poles in the scattering $T$-matrix. We
concentrate in the exotic channels with negative charm quantum number
for which there is the experimental claim of one state.
\end{abstract}

\pacs{}

\maketitle


\section{Introduction}

To understand the structure of mesons and baryons has been an active
topic of research in recent years.  One reason for that is the
numerous observations of states that do not fit the usual
interpretation of mesons as $q\bar q$ or baryons as $qqq$
states. Moreover the prospect of many new experiments (like FAIR at
GSI or the LHC) opens the possibility that even more resonances are to
be observed in the next years. Among the newly observed states many
new charmed baryon resonances have been found in the last few years
\cite{exp1,exp2,exp3,exp4,exp5}.  There has been in the past claims
for narrow baryon states which would be made of hidden charm
\cite{hid1,hid2} and also a claim of an exotic anti-charmed baryon
\cite{pentaq}.

From the theoretical point of view there has been a lot of success in
describing many baryon resonances as dynamically generated states in
coupled channels. For instance in \cite{lutz,estela} a rich spectrum
of charmed baryons is generated dynamically and accommodates many of
the experimentally observed states. Moreover, predictions are made in
\cite{hofmann} for the spectrum of hidden charm, anti-charmed, charmed
and $C=2$ $J^P=\oh^-$ baryons.  In all these cases, the states are
dynamically generated from the interaction of pseudo-scalar mesons
with ground state $J^P = \frac{1}{2}^+$ baryons. In \cite{hofmann2}
the interaction of the pseudo-scalar mesons with the ground state
$J^P=\trh^+$ baryons is studied, also giving rise to many dynamically
generated resonances. All these works consider for the interaction
Lagrangians based on chiral and SU(4) flavor symmetry and the KSFR
relation. The dynamics behind the interaction is assumed to be the
exchange of vector mesons in the Weinberg-Tomozawa term, while the
flavor symmetry is broken by using physical hadron masses. The
interaction of $\db N$ has been throughly studied in
\cite{haidenbauer} also assuming vector meson exchanges and
$\rm{SU}(4)$ constrains.

In this work we study the possibility of generating dynamically exotic
resonances with negative charm quantum number.  We follow an
approach~\cite{juan1} consistent with Heavy Quark Symmetry (HQS). HQS
predicts that all types of spin interactions vanish for infinitely
massive quarks: the dynamics is unchanged under arbitrary
transformations in the spin of the heavy quark. As a consequence, the
scheme of Ref.~\cite{juan1} takes on an equal footing the heavy vector
and pseudo-scalar mesons.

We briefly present in the next section the theoretical framework used
in the model. In Sec.  \ref{sec:res} we present and discuss on the
dynamically generated spectrum and in Sec. \ref{sec:concl} we summarize our
conclusions. Finally, there is an appendix where we collect the
different coupled channel matrices used in this work.


\section{Framework}\label{sec:frame}

We follow here the $\rm{SU}(8)$ spin flavor scheme of reference
\cite{juan1}.  This model is an extension of models based on
$\rm{SU}(4)$ flavor symmetry with $t$-channel vector meson exchanges
to an $\rm{SU}(8)$ spin-flavor scheme. Previously used models based on
$\rm{SU}(4)$ symmetry suffer from the limitation that they do not
include heavy pseudo-scalar and heavy vector mesons on an equal
footing. This is not justified from the point of view of HQS, which is
the proper spin-flavor symmetry of QCD if one takes the limit of
infinitely heavy quark masses.

In $\rm{SU}(8)$, the lowest lying baryons are represented by a
120-plet. To properly identify the spin-$1/2$ and spin-$3/2$ baryons
with the states of this multiplet, let us first decompose the 120-plet
into its inner $\rm{SU}(2)\otimes\rm{SU}(4)$ structure:
\be
120&\rightarrow& 20_2\oplus20^\prime_4.
\ee
The two 20-plets ($20_2$) of $\rm{SU}(4)$ can now accommodate the
spin-$1/2$ baryons while the four 20$^\prime$-plets represent the
spin-$3/2$ baryons. On the other hand the lowest lying mesons in
$\rm{SU}(8)$ are represented by a 63-plet that decomposes itself in
$\rm{SU}(2)\otimes\rm{SU}(4)$ as:
\be
63&\rightarrow& 15_1\oplus 15_3\oplus 1_3.
\ee
Note, however that the low lying mesons are constructed by the product
of an 8-plet of quarks with its conjugate:
\be
8\otimes\bar8\ &=&63\oplus 1.
\ee
This extra $1$ is assigned as the extra pseudo-scalar needed to form
the $\eta$, $\eta^\prime$ and $\eta_c$ mesons.

There are four possibilities to construct mesonic and baryonic
hadronic currents that can couple to a singlet in order to construct a
Lagrangian invariant under SU(8) rotations, but only one of these
possibilities reproduces the $\rm{SU}(3)$ Weinberg-Tomozawa Lagrangian
for the light mesons and baryons \cite{juan1}:
\be
\lc_{WT}^{SU(8)}\propto 
((M^\dagger\otimes M)_{63_a}\otimes(B^\dagger \otimes B)_{63})_1 .
\label{eq:wtlag}
\ee
The reduction of this Lagrangian to $\rm{SU}(6)$ reproduces the
Weinberg-Tomozawa Lagrangian used in \cite{recio,juan2,Toki:2007ab}.

The model has two main flavor symmetry breaking sources, first by the
use of physical masses for all the mesons and baryons and second by
the use of different meson decay constants. The values for the decay
constants of the mesons we use are:
\be
f_{D_s}=193.7~ {\rm MeV,} \quad f_D=f_{D^*}=f_{D_s^*}=157.4~ {\rm MeV}.
\ee
All meson-baryon pairs with the same $CSIJ$ quantum numbers (Charm,
Strangeness, Isospin and total angular momentum) span a coupled
channel space. The $s$-wave tree level amplitudes between two
channels for each $CSIJ$ sector is given by:
\be
V_{ij}^{CSIJ}&=&\xi_{ij}^{CSIJ}\frac{2\sqrt{s}-M_i-M_j}{4f_if_j} \nn \\
&\times& \sqrt{\frac{E_i+M_i}{2M_i}}\sqrt{\frac{E_j+M_j}{2M_j}}, \label{eq:pot}
\ee
where $\sqrt{s}$ is the center of mass energy of the system, $M_i$ is
the mass of the baryon in the $i^{th}$-channel, $E_i$ is the energy of
the C.M. baryon in the $i^{th}$-channel, $f_i$ is the decay
constant of the meson in the $i^{th}$-channel and $\xi_{ij}^{CSIJ}$
are coefficients coming from the $\rm{SU}(8)$ group structure of the
couplings. Tables for the $\xi$ coefficients can be found in the
appendix.

 We use this matrix $V$ as kernel to calculate the $T$-matrix:
\be
T^{CSIJ}&=&(1-V^{CSIJ}G^{CSIJ})^{-1}V^{CSIJ},
\ee
where $G^{CSIJ}$ is a diagonal matrix containing the two particle
propagators for each channel. Explicit expressions for the loop
functions can be found in the appendix of \cite{juan3} for the
different possible Riemann sheets.

The loop function diverges logarithmically and therefore must be
regularized. We choose to regularize it by a subtraction constant such
that~\cite{hofmann, hofmann2} 
\be
G_{ii}^{CSIJ}(\sqrt{s}=\mu^{CSIJ})=0,
\ee
and we choose $\mu^{CSIJ}$ to be $\sqrt{m_{\rm{th}}^2+M_{\rm{th}}^2}$
where $m_{\rm{th}}$ and $M_{\rm{th}}$ are respectively the masses of
the meson and baryon in the lowest threshold in the sector $CSIJ$. 

With all these ingredients we look now for poles of the
$T$-matrix. Poles on the first Riemann sheet below threshold are
interpreted as bound states.  Poles appearing in the second Riemann
sheet of open channels are interpreted as resonances and poles on the
second Riemann sheet, over the real axis but for closed channels are
interpreted as virtual states. Poles appearing in different positions
than the ones mentioned can not be associated with physical states and
are, therefore, artifacts that we call non-physical states.

For bound states and resonances the real part of the pole position is
associated with its mass and the imaginary part of the poles
interpreted as resonances is associated with one half of its
width. Other information that one can extract from the poles of the
$T$-matrix are the couplings of the states to their coupled
channels. Close to a pole the $T$-matrix can be written as:
\be
T^{CSIJ}_{ij}(z)&=&\frac{g_ig_j}{z-z_{pole}}, 
\ee
where $z_{pole}$ is the pole position in the $\sqrt{s}$ plane and the
$g_k$ is the dimensionless coupling of the resonance to channel $k$. So, by
calculating the residues of the $T$-matrix at some pole, one obtains
the product of the couplings $g_ig_j$.


\section{Results}\label{sec:res}

First we analyze the underlying $\rm{SU}(3)$ structure of the
interaction. The $J^P=\oh^+$ 20-plet of baryons is composed by four
$\rm{SU}(3)$ multiplets: an octet with $C=0$, which is identified with
the low lying octet of baryons to which the proton and the neutron
belong, an anti-triplet, a sextet and a triplet. The anti-triplet and
the sextet have $C=1$ quantum number. The anti-triplet is made by a
$S=0$ isospin singlet ($\Lambda_c$) and an $S=-1$ doublet ($\Xi_c$)
while the sextet is composed by an $S=-2$ singlet ($\Omega_c$), a
$S=-1$ doublet ($\Xi^\prime_c$) and an $S=0$ triplet
($\Sigma_c$). Finally there is the triplet with $C=2$ to which a
$S=-1$ singlet ($\Omega_{cc}$) and an $S=0$ doublet ($\Xi_{cc}$)
belong.
\be
20_{SU(4)}&\rightarrow& \left( \begin{array}{c}
                          3_{cc} \\
                          \bar{3}_c\oplus 6_c \\
                          8 \end{array} \right)_{SU(3)}.
\ee
The $J^P=\trh^+$ 20$^\prime$-plet of baryons is composed also by four
$\rm{SU}(3)$ multiplets: a decuplet with $C=0$ to which the low lying
$I=\trh$ $\Delta$ belongs, a sextet with $C=1$, a triplet with $C=2$
and a singlet with $C=3$.
\be
20^\prime_{SU(4)}&\rightarrow& \left( \begin{array}{c}
                                      1^*_{ccc} \\
                                      3^*_{cc} \\
                                      6^*_{c} \\
                                      10^* \end{array} \right)_{SU(3)}.
\ee
The $^*$ in a baryon multiplet indicates it is a $J^P=\trh^+$ baryon,
and we use a bar to denote the conjugate representations.

The pseudo-scalar and vector mesons have similar structure, they
belong to 15-plets of $\rm{SU}(4)$. These 15-plets break down into
four $\rm{SU}(3)$ multiplets namely, a triplet, an octet, a singlet
and an anti-triplet. The octets have null charm quantum number and are
identified with the low lying pseudo-scalar and vector mesons ($\pi$,
$K$, $\eta$, $\rho$, $K^*$ and $\omega$).  We use pure $\bar c c$ wave
functions for lowest charmonium states $\eta_c$ and $J/\psi$ and mix
with the physical $\eta, \eta'$ and $\omega, \phi$ mesons,
respectively, to build the charmless singlet present in the SU(4) 15
plet.  The anti-triplets are identified with the $D$ doublets and the
$D_s$. The triplets are the antiparticles of the anti-triplets.
\be
15_{SU(4)}&\rightarrow& \left( \begin{array}{c}
                                \bar{3}_c \\
                                8\oplus1 \\
                                3_{\bar{c}} \end{array} \right)_{SU(3)}.
\ee
To differentiate the pseudo-scalars and the vector mesons we write a
$^*$ after the number indicating the vector multiplet.

The only way to have meson baryon systems with $C=-1$ is to couple a
$C=0$ baryon with a $C=-1$ meson. Considering the angular momentum of
the particles we have the following options\footnote{The baryon
multiplet comes first in the irrep products.}:
\begin{itemize}
\item For $J=\oh$ \\ \\
  $8\otimes3_{\bar{c}} = 3\oplus \bar{6}\oplus 15$\\ \\
  $8\otimes3^*_{\bar{c}} = 3\oplus \bar{6}\oplus 15$\\ \\
  $10^*\otimes3^*_{\bar{c}} = 15\oplus 15^\prime$\\
\item For $J=\trh$\\ \\
  $8\otimes3^*_{\bar{c}} = 3\oplus \bar{6}\oplus 15$\\ \\
  $10^*\otimes3_{\bar{c}} = 15\oplus 15^\prime$\\ \\
  $10^*\otimes3^*_{\bar{c}} = 15\oplus 15^\prime$\\ \\
\item For $J=\frac{5}{2}$\\ \\
  $10^*\otimes3^*_{\bar{c}} = 15\oplus 15^\prime$
\end{itemize}

The study of the eigenvalues of the $\xi$ matrices in \Eq{eq:pot} for
the different sectors indicates whether a given multiplet is
attractive or repulsive and therefore in which sectors and how many
poles one expects to generate \cite{rocaev,meusca}.  For $J=\oh$ the
two triplets, one $\bar{6}$-plet, two 15-plets and the
$15^\prime$-plet are attractive. In the $J=\trh$ sector the triplet,
the $\bar{6}$-plet, two 15-plets and one $15^\prime$-plet are
attractive and in the $J=\frac{5}{2}$ only the 15-plet is attractive.

In order to track down each pole to a definite $\rm{SU}(3)$ multiplet
we start from an $\rm{SU}(3)$ symmetric scheme by setting the masses of
all particles belonging to the same $\rm{SU}(3)$ multiplet to a common
value. In this $\rm{SU}(3)$ limit we use the following values for the
masses of the mesons, which are approximately the average value of the
mass\footnote{For the physical masses of the mesons we use,
$m_D$=1.867 GeV, $m_{D_s}$=1.968 GeV, $m_{D^*}$=2.0085 GeV and
$m_{D_s^*}$=2.112 GeV and for the physical masses of the baryons we
use, $m_N$=0.939 GeV, $m_\Lambda$=1.116 GeV, $m_\Sigma$=1.193 GeV,
$m_\Xi$=1.318 GeV, $m_\Delta$=1.210 GeV, $m_{\Sigma^*}$= 1.385 GeV,
$m_{\Xi^*}$=1.533 GeV and $m_\Omega$=1.672 GeV.} in each multiplet,
$m_{3_{\bar c}}$=1.9 GeV and $m_{3_{\bar c}^*}$=2.05 GeV and for the
baryons, $m_8$=1 GeV and $m_{10^*}$=1.4 GeV.

To gradually break $\rm{SU}(3)$ symmetry we write the mass of the
hadrons as a function of a parameter $x$ such that
\be
m(x)=\bar{m}+x(m_{\rm{phys}}-\bar{m})\label{eq:su3bre},
\ee
where $\bar{m}$ is the mass of the hadron in the $\rm{SU}(3)$ limit
and $m_{\rm{phys}}$ is the physical mass of the particle. In this way,
we vary $x$ between 0 and 1, 0 being the $\rm{SU}(3)$ limit
and 1 the real world. We also change $f_{D_s}$ to gradually approach
the value of $f_D$ when restoring the $\rm{SU}(3)$ symmetry.

We show in Tables \ref{tab:poles1}, \ref{tab:poles2} and
\ref{tab:poles3} the pole positions we find for the $J^P=\oh^-$,
$J^P=\trh^-$ and $J^P=\frac{5}{2}^-$ sectors, respectively. We also show
 in the tables the two channels to which each resonance has the
strongest couplings, those should be the most important components in
the wave function of each resonance \cite{meuwave}. 

Some of the resonances are bound by energies of the order of 200-300
MeV in relation to the thresholds of their main channels. Our approach
is based in the Weinberg-Tomozawa term of \Eq{eq:wtlag}. This
Lagrangian is roughly the first order term in a low momentum
expansion. The deeper the bound states are, the more relevant higher
absolute values of the momentum (because of phase space)
become. Therefore, we expect theoretical uncertainties affecting to
our results for such states would be bigger, since higher order
Lagrangians should give sizable corrections.

On the other hand some of the states which we obtain are bound by 150
MeV or less. Our results for such states are expected to be more
precise, and we will focus on these states in the next subsections.

\begin{table}
\begin{center}
\caption{Pole positions in the $J^P=\oh^-$ sector. In the column SU(3)
Irrep / pole, we show the pole position in the $\rm{SU}(3)$ symmetric
limit. The (*) indicates a non-physical pole (placed in the second
Riemann sheet below threshold). The column Main Channels shows the
channels to which the resonance couples more strongly and in the
$B_1~ \& ~ B_2$ column, we show the energy difference between the threshold of
each of these channels and the resonance mass.} \label{tab:poles1} 

\vspace{0.4cm}

\begin{tabular}{c||c|c|c|c|c|c}
\hline\tstrut
 SU(3) Irrep      &$S$&$I$& Re($\sqrt{s}$)& Im($\sqrt{s}$) & Main & $B_1$ \& $B_2$ \\
 Pole [MeV] & &         &     [MeV]  &  [MeV]     & Channels &  [MeV]   \\
\hline
\hline
   &     &          &    &  & & \\
           &  0  &  2      &3125.7&0 & $\del\dvb$ & 93\\
\cline{2-3}\tstrut
$15^\prime$& $-$1  & $3/2$  &3208.5&$-$0.5& $\sigs \dvb$, $\del\dsvb$ & 185  \& 113 \\
\cline{2-3}\tstrut 
         &   $-$2  &  1      &3309.9&$-$0.3& $\css\dvb$, $\sigs\dsvb$ & 231 \& 187\\
\cline{2-3}\tstrut
3242.7   &   $-$3  & $1/2$   &3402.0&$-$0.2& $\css\dsvb$, $\Omega\dvb$ & 243 \& 278\\
\cline{2-3}\tstrut
         &   $-$4  & 0       &3543.2&0 & $\Omega\dsvb$ & 241\\
\hline
   &     &          &    &  & & \\
         &    0  & 1       &2872.6&$-$45.6 & $N\dvb$, $\del\dvb$ & 75 \& 346\\
\cline{2-3}\tstrut
15       &   $-$1  & $1/2$   &2995.8&$-$3.6& $\Sigma\db$, $\Lambda\dvb$ & 65 \& 129\\
\cline{3-3}\tstrut
         &       & $3/2$  &3048.7&0& $\Sigma\dvb$, $\Sigma\db$ & 153 \& 12\\
\cline{2-3}\tstrut
2945.7   &   $-$2  & 0       &3109.0&$-$12.0& $\Lambda\dsvb$, $\Xi\dv$ & 119 \& 76\\
\cline{3-3}\tstrut
-i\ 36.4  &       & 1       &3160.9&0&  $\Xi\dvb$, $\Sigma\dsb$ & 166 \& $<$1\\
\cline{2-3}\tstrut
         &   $-$3  & $1/2$   &3267.2&0& $\Xi\dsvb$, $\Xi\dsb$ & 163 \& 19\\
\hline
   &     &          &    &  & & \\
         &    0  & 1       &3002.0&$-$52.4& $\del\dvb$, $N\db$ & 216 \& $-$195\\
\cline{2-3}\tstrut
15       &   $-$1  & $1/2$   &3104.1&$-$27.5& $\sigs\dvb$, $\Lambda\dvb$ & 289 \& 20\\
\cline{3-3}\tstrut
         &       & $3/2$  &3135.0&$-$5.3& $\del\dsvb$, $\Sigma\dvb$ & 187 \& 66\\
\cline{2-3}\tstrut
3124.5   &   $-$2  & 0       &3216.4&$-$21.2& $\css\dvb$, $\Xi\dvb$ & 325 \& 110\\
\cline{3-3}\tstrut
-i\ 57.8  &       & 1       &3239.2&$-$12.4& $\sigs\dsvb$, $\css\dvb$ & 258 \& 302\\
\cline{2-3}\tstrut
         &   $-$3  & $1/2$   &3337.0 &$-$16.6&  $\Omega\dvb$, $\css\dsvb$ & 343 \& 308\\
\hline
   &     &          &    &  & & \\
         &   0   & 0       &2805.0&0& $N\db$, $N\dvb$ & 1 \& 142\\
\cline{2-3}\tstrut
$\bar{6}$&   $-$1  & $1/2$   &2971.7&$-$3.1& $N\dsvb$, $\Lambda\db$ & 79 \& 12\\
\cline{2-3}\tstrut
         &   $-$2  & 1       &3126.0&0& $\Xi\db$, $\Sigma\dsvb$ & 59 \& 179\\
2890.2   &     &          &      & & &\\
\hline
   &     &          &    &  & & \\
3        &   $-$1  & $1/2$   &2861.0(*)&$-$74.2& $N\dsb$, $\Lambda\dvb$ & 46 \& 263\\
\cline{2-3}\tstrut
         &   $-$2  & 0       &3080.1&0& $\Xi\db$, $\Lambda\dsb$ & 105 \& 4\\
2868.9  &      &        &       &  & &\\
\hline
   &     &          &      & & \\
3        &   $-$1  & $1/2$   &3049.3&$-$8.2& $\Sigma\dvb$, $N\dsvb$ &152 \& 2\\
\cline{2-3}\tstrut
         &   $-$2  & 0       &3169.2&$-$6.7& $\Xi\dvb$, $\Lambda\dsvb$ & 157 \& 59\\
2994.0   &      &          &       & & &\\    
\hline
\end{tabular}
\end{center}
\end{table}

\begin{table}
\begin{center}
\caption{Same as Table \ref{tab:poles1} for the $J^P=\trh^-$ sector. } \label{tab:poles2}

\vspace{0.4cm}

\begin{tabular}{c||c|c|c|c|c|c}
\hline\tstrut
 SU(3) Irrep      &$S$&$I$& Re($\sqrt{s}$)& Im($\sqrt{s}$) & Main & $B_1$ \& $B_2$ \\
 Pole [MeV] & &         &     [MeV]  &  [MeV]     & Channels &  [MeV]   \\
\hline
\hline
   &     &          &     & & & \\
           &  0  &  2      &3061.1&0& $\del\dvb$, $\del\db$ & 157 \& 16 \\
\cline{2-3}\tstrut
$15^\prime$& $-$1  & $3/2$  &3189.7&$-$5.6& $\sigs\dvb$, $\sigs\db$ & 204 \& 63\\
\cline{2-3}\tstrut
         &   $-$2  &  1      &3326.5&$-$0.1& $\css\dvb$, $\sigs\dsb$ & 215 \& 26\\
\cline{2-3}\tstrut
3221.5   &   $-$3  & $1/2$   &3433.8&$<$0.1& $\Omega\dvb$, $\Xi\dsvb$ & 247 \& 4\\
\cline{2-3}\tstrut
         &   $-$4  & 0       &3533.4&0& $\Omega\dsvb$, $\Omega\dsb$ & 251 \& 107\\
\hline
   &     &          &   &   & & \\
         &    0  & 1       &2978.6&$-$16.0& $\del\db$, $N\dvb$ & 99 \& $-$31\\
\cline{2-3}\tstrut
15       &   $-$1  & $1/2$   &3116.6&$-$0.6& $\Sigma\dvb$, $\sigs\db$ & 84 \& 136\\
\cline{3-3}\tstrut
         &       & $3/2$  &3151.8&0& $\del\dsb$, $\sigs\db$ & 26 \& 101\\
\cline{2-3}\tstrut
3115.6   &   $-$2  & 0       &3224.1&0& $\css\db$, $\Xi\dvb$ & 176 \& 102\\
\cline{3-3}\tstrut
-i\ 29.3   &       & 1       &3260.9&0& $\css\db$, $\css\dvb$ & 140 \& 281\\
\cline{2-3}\tstrut
         &   $-$3  & $1/2$   &3346.3&0& $\Xi\dsvb$, $\Omega\db$ & 84 \& 193\\
\hline
   &     &          &    &  & & \\
         &    0  & 1       &3066.8&$-$16.8& $\del\dvb$, $N\dvb$ & 152 \& $-$119\\
\cline{2-3}\tstrut
15       &   $-$1  & $1/2$   &3190.2&$-$9.3& $\sigs\dvb$, $\sigs\db$ & 203 \& 62\\
\cline{3-3}\tstrut
         &       & $3/2$  &3175.8&0& $\del\dsvb$, $\sigs\db$ & 146 \& 77\\
\cline{2-3}\tstrut
3216.4   &   $-$2  & 0       &3296.1&$-$4.6& $\css\dvb$, $\css\db$ & 245 \& 104\\
\cline{3-3}\tstrut
-i \ 11.8   &       & 1       &3295.9&0& $\sigs\dsvb$, $\css\db$ & 201 \& 105\\
\cline{2-3}\tstrut
         &   $-$3  & $1/2$   &3397.3&0& $\css\dsvb$, $\Omega\dvb$ & 248 \& 283\\
\hline
   &     &          &    &  & & \\
         &   0   & 0       &2922.1&0& $N\dvb$ & 25\\
\cline{2-3}\tstrut
$\bar{6}$&   $-$1  & $1/2$   &3029.7&0& $N\dsvb$, $\Lambda\dvb$ & 21 \& 95\\
\cline{2-3}\tstrut
         &   $-$2  & 1       &3206.2&0& $\Sigma\dsvb$, $\Xi\dvb$ & 99 \& 120\\
3008.1   &     &          &   &   & & \\
\hline
   &     &          &    &  & & \\
3        &   $-$1  & $1/2$   &3097.5&$-$1.8& $\Sigma\dvb$, $\sigs\db$ & 104 \& 155\\
\cline{2-3}\tstrut
         &   $-$2  & 0       &3181.6&0& $\Xi\dvb$, $\Lambda\dsvb$ & 145 \& 46\\
3008.1   &      &      & &    &       &\\    
\hline
\end{tabular}
\end{center}
\end{table}

\begin{table}
\begin{center}
\caption{Same as Table \ref{tab:poles1} for the $J^P=\frac52^-$ sector.} \label{tab:poles3}

\vspace{0.4cm}

\begin{tabular}{c||c|c|c|c|c|c}
\hline\tstrut
 SU(3) Irrep      &$S$&$I$& Re($\sqrt{s}$)& Im($\sqrt{s}$) & Main & $B_1$ \& $B_2$ \\
 Pole [MeV] & &         &     [MeV]  &  [MeV]     & Channels &  [MeV]   \\

\hline
\hline
   &     &          &    &  & & \\
         &    0  & 1       &3125.7&0& $\del\dvb$ & 93\\
\cline{2-3}\tstrut
15       &   $-$1  & $1/2$   &3252.4&0& $\sigs\dvb$ & 141\\
\cline{3-3}\tstrut
         &       & $3/2$  &3224.6&0& $\del\dsvb$, $\sigs\dvb$ & 97 \& 169\\
\cline{2-3}\tstrut
3298.2   &   $-$2  & 0       &3444.4&0& $\css\dvb$ & 201\\
\cline{3-3}\tstrut
         &       & 1       &3349.0&0& $\css\dvb$, $\sigs\dsvb$ & 192 \& 148\\
\cline{2-3}\tstrut
         &   $-$3  & $1/2$   &3455.7&0& $\Omega\dvb$, $\css\dsvb$ & 225 \& 189\\
\hline
\end{tabular}
\end{center}
\end{table}


\subsection{The $J^P=\frac{1}{2}^-$ states}

The $J^P=\oh^-$ states are shown in Table \ref{tab:poles1}. With
exception of the resonance in the $S=0$, $I=2$ sector, all of the
resonances coming from the 15$^\prime$-plet are too bound for our
model to give precise numerical predictions. This happens also for the
resonances coming from one of the 15-plets. This is due to the fact
that for these $15$ and $15^\prime$ multiplets, the poles are
generated from the interaction between the $10^*$ (baryon decuplet)
with the $3^*$ (anti-charmed vector mesons). And this is different
from the other $15$-plet that in the $\rm{SU}(3)$ limit appears for
$\sqrt{s}=(2945.7-i36.4)$ MeV, where the poles come basically from the coupling
$8\otimes 3^*$ ($J^P=\oh^+$ baryon decuplet with the vector meson triplet).
So, the two 15-plets have different structures.

Lighter and less bound resonances are the ones belonging to the two
triplets and to the anti-sextet. Among these resonances we call the
attention to the $S=0$, $I=0$ member of the sextet, generated by the
$N\db$ and $N\dvb$ coupled channel dynamics. This state is bound by
only 1 MeV, and it is one of our more interesting
predictions. Moreover, it appears as a consequence of treating heavy
pseudo-scalars and heavy vector mesons on an equal footing, as
required by HQS. Indeed, if one looks at the coupled channel matrix,
$\xi_{ij}$, in Table \ref{tab:tab1mev}, one finds the diagonal $N\db
\to N\db$ entry is zero, which means that no interaction in this
sector would be generated if the $N\dvb$ channel is not considered, as
it was the case in Ref. \cite{hofmann}. However, the inclusion of this
latter channel gives rise to an attractive eigenvalue ($\lambda=-2$)
for the eigenvector: $ \frac{\sqrt{3}}{2}\ |N\db \rangle + \frac12 \
|N \dvb\rangle $, which originates the bound state reported in Table
\ref{tab:poles1}.

\begin{figure}
\begin{center}
\includegraphics[width=8.5cm,angle=-0]{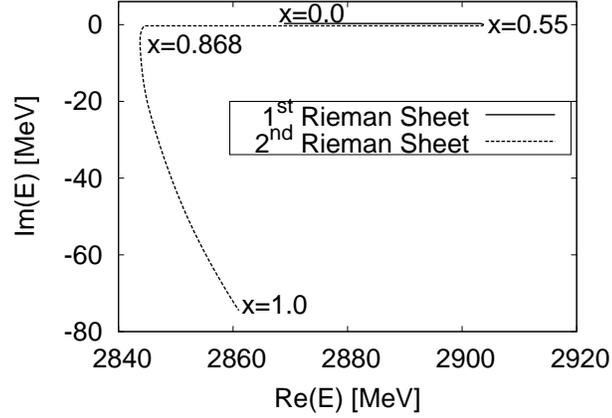}
\caption{Path followed by the $J^P=\oh^-, S=-1$, $I=\oh$ pole marked
with an $*$ in Table~\ref{tab:poles1} while changing the $x$ parameter
between 0 and 1. Around $x=0.55$ the pole goes from the first to the
second Riemann sheet and around $x=0.868$ it starts moving away from
the real axis.}  \label{fig:pole}
\end{center}
\end{figure}

The breaking of the SU(3) symmetry through the parameter x of \Eq{eq:su3bre} makes the $S=-1$,
$I=\oh$ pole, which is member of the lightest of the triplets,
first approach the $N \bar D_s$ threshold up to x=0.55 but, instead of crossing it at
some point and becoming a resonance able to decay into this channel, it turns into a
virtual state. That is, it moves into the second Riemann sheet, and stays below threshold moving
away from it over the real axis. However, for values of x bigger than 0.868, it starts
acquiring an imaginary part and can no longer be interpreted as a physical state. This
behaviour is shown in Fig. \ref{fig:pole}, where we present a plot of the path that this pole
describes while breaking the SU(3) symmetry.

This pole and its $S=-2$ companion had already been predicted in
\cite{hofmann}. In that work only the interaction of pseudo-scalars
and $J^P=\oh^+$ baryons were considered and the states obtained there
turn out to be much more bound (binding energies of about 130-250 MeV)
than in the present approach. In this case, vector meson degrees of
freedom play a minor role, and the origin of the discrepancy now
should be traced back to our pattern of SU(4) flavour breaking, which
makes our interaction weaker in this sector by a factor $(f_\pi/f_D)^2$ than that used in
\cite{hofmann}. For the sextet
there is also predictions of pentaquarks based in the Skyrme model in
\cite{skyrme}. The masses of the states in that paper are 100 MeV
lower than the ones we find with our approach.
We also show in Tables \ref{tabcoup61}, \ref{tabcoup31} and \ref{tabcoup32}
the couplings of the members of the anti-sextet and the two triplets
to the different channels that appear in their dynamics.

\begin{table}
\begin{center}
\caption{Couplings of the poles belonging to the anti-sextet of
Table~\ref{tab:poles1}.} \label{tabcoup61}
\vspace{0.2cm}
\begin{tabular}{ccc}
\hline\tstrut
$S=0$, $I=0$~~~~ & $S=-1$, $I=1/2$~~~~ & $S=-2$, $I=1$  \\
\hline
\hline\tstrut
& & \\
\begin{tabular}{c|c}
\hline\tstrut
Channel & $|g_i|$ \\
\hline\tstrut
& \\
$N\db$ & 1.5 \\
$N\dvb$ & 1.4 \\
\hline
\end{tabular}  &
                      \begin{tabular}{c|c}
                      \hline\tstrut
                      Channel & $|g_i|$ \\
                      \hline\tstrut
                      & \\
                      $N\dsb$ & 0.7 \\
                      $\Lambda\db$ & 2.0 \\
                      $N\dsvb$ & 2.5 \\
                      $\Sigma\db$ & 1.2 \\
                      $\Lambda\dvb$ & 0.5 \\
                      $\Sigma\dvb$ & 0.8 \\
                      $\sigs\dvb$ & 1.0 \\
                      \hline
                      \end{tabular} &
                                              \begin{tabular}{c|c}
                                              \hline\tstrut
                                              Channel & $|g_i|$ \\
                                              \hline\tstrut
                                              & \\
                                              $\Sigma\dsb$ & 1.1 \\
                                              $\Xi\db$ & 2.7 \\
                                              $\Sigma\dsvb$ & 2.6 \\
                                              $\Xi\dvb$ & 0.3 \\
                                              $\sigs\dsvb$ & 0.7 \\
                                              $\css\dvb$ & 0.6 \\
                                              \hline
                                              \end{tabular} \\

\end{tabular}
\end{center}
\end{table}
\begin{table}
\begin{center}
\caption{ Couplings of the physical pole belonging to the lightest
of the triplets of Table \ref{tab:poles1}.} \label{tabcoup31}
\vspace{0.2cm}
\begin{tabular}{c}
\hline\tstrut
$S=-2$, $I=0$  \\
\hline
\hline\tstrut
\\
\begin{tabular}{c|c}
\hline
Channel & $|g_i|$ \\
\hline
& \\
$\Lambda\dsb$ & 1.4 \\
$\Xi\db$ & 2.8 \\
$\Lambda\dsvb$ & 0.1 \\
$\Xi\dvb$ & $<$0.1 \\
$\css\dvb$ & $<$0.1 \\
\hline
\end{tabular}  \

\end{tabular}
\end{center}
\end{table}

\begin{table}
\begin{center}
\caption{Couplings of the poles 
belonging to the heaviest of the triplets of Table 
\ref{tab:poles1}.} \label{tabcoup32}
\vspace{0.2cm}
\begin{tabular}{cc}
\hline\tstrut
$S=-1$, $I=1/2$ ~~~~& $S=-2$, $I=0$  \\
\hline
\hline\tstrut
\\
\begin{tabular}{c|c}
\hline\tstrut
Channel & $|g_i|$ \\
\hline\tstrut
& \\
$N\dsb$ & 0.5 \\
$\Lambda\db$ & 0.9 \\
$N\dsvb$ & 1.8 \\
$\Sigma\db$ & 0.1 \\
$\Lambda\dvb$ & 1.2 \\
$\Sigma\dvb$ & 3.2 \\
$\sigs\dvb$ & 1.7 \\
\hline
\end{tabular}  &
                      \begin{tabular}{c|c}
                      \hline
                      Channel & $|g_i|$ \\
                      \hline
                      & \\
                      $\Lambda\dsb$ & 0.9 \\
                      $\Xi\db$ & 0.6 \\
                      $\Lambda\dsvb$ & 2.0 \\
                      $\Xi\dvb$ & 4.1 \\
                      $\css\dvb$ & 1.6 \\
                      \hline
                      \end{tabular} \\

\end{tabular}
\end{center}
\end{table}


\subsection{The $J^P=\frac{3}{2}^-$ states}

The $J^P=\trh^-$ states are shown in Table~\ref{tab:poles2}.  The
situation here is similar to the case of the $J^P=\oh^-$ states. The
states in the 15$^\prime$-plet and the 15-plets are strongly bound
while the poles in the anti-sextet and the triplet turn out to be less
bound. In any case one should notice that one of the members of the
heaviest 15-plet is a candidate for the claimed anti-charmed pentaquark in
\cite{pentaq}.  The pentaquark was claimed to be seen in $D^{*-}p$ and
$D^{*+}\bar{p}$ spectrum with a mass $m=3099$ MeV.  There is, though,
controversy about the existence of this state since other experiments
failed to confirm it \cite{other}.  The $S=0$, $I=1$ pole at position
$\sqrt{s}=(3066.8-i16.8)$ MeV is basically a bound state of $\del\dvb$
and has sizable couplings to $\del\db$ and $N\dvb$. The $N\dvb$
channel is open and is responsible for the $\sim 30$ MeV width of the
resonance. The state claimed in \cite{pentaq} has been observed in the
decay mode
\be
\Theta_{\bar C}&\rightarrow& N \dvb \rightarrow N \db\pi, \label{eq:dec1}
\ee
where the $\dvb$ has been identified from its decay first to a soft
pion with a $\bar D^0$ meson and the subsequent decay of the $\bar
D^0$ to a $K^-\pi^+$. Our dynamically generated state has other two
possible decay channels induced by its coupling to channels involving
the $\del (1232)$ resonance, which is not a stable particle. Thus, the
anti-charmed resonance can decay to $\db$ or $\dvb$ plus a virtual
$\del$, which subsequently would decay into a $\pi N $ pair.  To take
this possibility into account, we follow the prescription used in
\cite{roca,meuax} and convolute the loop function of the channels with
the $\del$ with the spectral function of this resonance:
\be
\tilde{G}(\sqrt{s},m,M_\del)&=&{1\over N}\int_{(m_\del-2\Gamma_\del)^2}^{(m_\del+2\Gamma_\del)^2}d\tilde{M}^2{\cal S}(\tilde{M}^2,m_\del,\Gamma_\del)\nn\\
&\times&  G(\sqrt{s},m,\tilde{M}) \\
N&=&\int_{(m_\del-2\Gamma_\del)^2}^{(m_\del+2\Gamma_\del)^2}d\tilde{M}^2{\cal S}(\tilde{M}^2,m_\del,\Gamma_\del)\nn \\
{\cal S}(\tilde{M}^2,m_\del,\Gamma_\del)&=&-{1\over\pi} \rm{Im}\ \left ( {1\over \tilde{M}^2-m_\del^2+im_\del\Gamma_\del}\right ).
\ee
We have also changed the subtraction point in order to get the mass of
this resonance closer to the claimed value of 3100 MeV.  We did that
by increasing 10\% the value of the subtraction point. The new pole
position, taking into account the 120 MeV width of the $\del$ and the
slightly changed subtraction point $\mu$ is $\sqrt{s}=(3098.2-i38.0)$
MeV. We note though that our resonance has a much bigger width ($\sim
70$ MeV) than the one observed experimentally ($12\pm3$ MeV).  However, our
dynamically generated state has now another two decay mechanisms apart
from the one in \Eq{eq:dec1}, namely
\be
\Theta_{\bar C}&\rightarrow&\del \db\rightarrow N\pi\db \label{eq:dec2}\\
\Theta_{\bar C}&\rightarrow&\del \dvb\rightarrow N\pi\db\pi.
\ee
The decay in \Eq{eq:dec2} has the same particles in the final state
that in \Eq{eq:dec1}, with the difference that the pion in one case is
coming from the decay of the $D^*$ and therefore has low momentum,
while in the other channel it comes from a $\del$ and may have higher
momentum. The experimental search made in \cite{pentaq} looked only
for pions in order to reconstruct a $D^*$ and may have missed the
other events where the pion comes from a $\del$.

Observing the Table~\ref{tab:poles2}, one actually sees that there are
other poles with $S=0$ that can decay to $N\dvb$, but their masses do
not agree well with that of the observed experimental state. On the
other hand, we see also that some other states from the table may be
affected by the consideration of the $\del$ or $\sigs$ widths. We show
in Table~\ref{tab:wwidth}, the new pole positions of the resonances
sizable affected by considering the widths of the $\del$
($\Gamma_\del$ =120 MeV) and $\sigs$ ($\Gamma_\sigs$=35 MeV). For
$J^P=\oh^-$ and $J^P=\frac{5}{2}^-$ only two states are notably
affected in each case but in the case of $J^P=\trh^-$ many more states
become broader because of the consideration of the decay of its
unstable components.
\begin{table}
\begin{center}
\caption{New pole positions for the resonances affected by the
consideration of the $\del$ and $\sigs$ widths.} \label{tab:wwidth}
\vspace{0.4cm}
\begin{tabular}{c||c|c|c|c}
\hline\tstrut
 SU(3) Irrep      &$S$&$I$& Re($\sqrt{s}$)& Im($\sqrt{s}$)   \\
Pole [MeV] & &         &     [MeV]  &  [MeV]   \\
\hline
\hline\tstrut
   &     &          &    &   \\
 $15^\prime$      &    0  & 2       &3114.9&$-$13.8 \\
\cline{2-3}\tstrut
$J^P=1/2^-$       &   $-$1  & 3/2   &3206.6&$-$2.4\\
\hline\tstrut
 $15^\prime$ & 0 & 2 & 3055.7 & $-$19.1 \\
\cline{2-3}\tstrut
$J^P=3/2^-$ & $-$1 & 3/2 & 3187.2 & $-$16.9 \\
\cline{2-3}\tstrut
   &$-$2 & 1 & 3324.1 & $-$2.5 \\
\hline
15 & 0 & 1 & 2964.0 & $-$26.0 \\
\cline{2-3}\tstrut
$J^P=3/2^-$ & $-$1 & 3/2 & 3156.1 & $-$19.6 \\
\hline\tstrut
15 & 0 & 1 & 3062.4 & $-$26.8 \\
\cline{2-3}\tstrut
$J^P=3/2^-$ & $-$1 & 3/2 & 3168.7 & $-$3.9 \\
\hline\tstrut
15 & 0 & 1 & 3114.9 & $-$13.8 \\
\cline{2-3}\tstrut
$J^P=5/2^-$ & $-1$ & 3/2 & 3216.0 & $-$9.9 \\
\hline
\end{tabular}
\end{center}
\end{table}
We can also compare here our states with previous predictions from
\cite{skyrme}. In that paper, with the Skyrme model, a 15-plet of
pentaquarks was found. The masses of the pentaquark states are between
100 and 200 MeV lower than the masses of the states belonging to the
lighter 15-plet that we generate dynamically. Predictions for a
negative charmed 15-plet had also been made in \cite{hofmann2}. In
that paper only the interaction of pseudo-scalar mesons with the
ground state of $J^P=\trh^+$ baryons have been considered, while in
the present work also the vector mesons and the $J^P=\oh^+$ baryons
are taken into account. The results obtained in that paper are to be
compared with the light 15-plet obtained in the present work, since
this multiplet is mainly coming from the $10^*\otimes 3$, as can be
seen from Table \ref{tab:coup15} where the couplings of the states to
the different channels are given. Also here the states obtained with
the $\rm{SU}(8)$ model are 100-200 MeV heavier (less bound) than in
the previous work~\cite{hofmann2}, which did not considered the heavy
quark symmetry. Part of this difference comes here again because our
interactions are weaker than those used in this latter reference by a
factor of the order of $(f_\pi/f_D)^2$.  The couplings of the
resonances can be also compared in both works. The strength of the
couplings in the present approach are mostly in the channels coming
from the $10^*\otimes 3$ interaction but there is a sizeable mixing
with states coming from the $10^*\otimes 3^*$ or $8\otimes 3^*$ in
some sectors. Moreover, since our states are less bound, the total
amount of strength in the couplings is smaller \cite{meuwave} than in
\cite{hofmann2}.

We also compile in Tables \ref{tab:coup6b} and \ref{tab:coup3} the
couplings of the resonances belonging to the anti-sextet and the
triplet from Table~\ref{tab:poles2}. The poles of the anti-sextet are
placed relatively close to threshold, where the current approach might
be more appropriate, and their dynamics is mostly controlled by channels
involving vector meson degrees of freedom not considered up to
now. On the other hand, they are sufficiently bound (tens of MeV) to
expect their existence to be guarantied under reasonable changes of
the employed renormalization scheme or of the pattern of SU(8) symmetry
breaking. Finally, in Table \ref{tab:coup15he} we present the couplings
of the poles belonging to the heaviest 15-plet of Table \ref{tab:poles2}.

\begin{table}
\begin{center}
\caption{Couplings of the poles belonging to the lightest 15-plet of
Table~\ref{tab:poles2}.} \label{tab:coup15}
\vspace{0.2cm}
\begin{tabular}{ccc}
\hline\tstrut
$S=0$, $I=1$ & $S=-1$, $I=1/2$ & $S=-1$, $I=3/2$  \\
\hline
\hline\tstrut
\\
\begin{tabular}{c|c}
\hline\tstrut
Channel & $|g_i|$ \\
\hline\tstrut
& \\
$N\dvb$ & 1.8 \\
$\del\db$ & 4.1 \\
$\del\dvb$ & 1.6 \\
\hline
\end{tabular}  &
                      \begin{tabular}{c|c}
                      \hline \tstrut
                      Channel & $|g_i|$ \\
                      \hline\tstrut
                      & \\
                      $N\dsvb$ & 0.3 \\
                      $\Lambda\dvb$ & 0.6 \\
                      $\Sigma\dvb$ & 3.0 \\
                      $\sigs\db$ & 2.8 \\
                      $\sigs\dvb$ & 0.6\\
                      \hline 
                      \end{tabular} &
                                              \begin{tabular}{c|c}
                                              \hline\tstrut
                                              Channel & $|g_i|$ \\
                                              \hline\tstrut
                                              & \\
                                              $\del\dsb$ & 2.2 \\
                                              $\Sigma\dvb$ & 1.6 \\
                                              $\sigs\db$ & 1.9 \\
                                              $\del\dsvb$ & 0.3 \\
                                              $\sigs\dvb$ & 1.2 \\
                                               \hline    
                                              \end{tabular} \\

\\
\hline\tstrut
$S=-2$, $I=0$ & $S=-2$, $I=1$ & $S=-3$, $I=1/2$  \\
\hline
\hline\tstrut
\\
\begin{tabular}{c|c}
\hline\tstrut
Channel & $|g_i|$ \\
\hline\tstrut
& \\
$\Lambda\dsvb$ & 0.6 \\
$\Xi\dvb$ & 2.3 \\
$\css\db$ & 3.5 \\
$\css\dvb$ & 0.8 \\
\hline
\end{tabular}  &
                      \begin{tabular}{c|c}
                      \hline
                      Channel & $|g_i|$ \\
                      \hline\tstrut
                      & \\
                      $\Sigma\dsvb$ & 1.1 \\
                      $\Xi\dvb$ & 1.3 \\
                      $\sigs\dsb$ & 1.6 \\
                      $\css\db$ & 2.9 \\
                      $\sigs\dsvb$ & 0.4 \\
                      $\css\dvb$ & 1.9 \\
                      \hline
                      \end{tabular} &
                                              \begin{tabular}{c|c}
                                              \hline\tstrut
                                              Channel & $|g_i|$ \\
                                              \hline\tstrut
                                              & \\
                                              $\Xi\dsvb$ & 1.9 \\
                                              $\css\dsb$ & 1.1 \\
                                              $\Omega\db$ & 3.5 \\
                                              $\css\dsvb$ & 0.6 \\
                                              $\Omega\dvb$ & 1.5 \\
                                              \hline
                                              \end{tabular} \\

\end{tabular}
\end{center}
\end{table}
\begin{table}
\begin{center}
\caption{Couplings of the poles belonging to the anti-sextet of
Table \ref{tab:poles2}.} \label{tab:coup6b}
\vspace{0.2cm}
\begin{tabular}{ccc}
\hline\tstrut
$S=0$, $I=0$~~~~& $S=-1$, $I=1/2$~~~~ & $S=-2$, $I=1$  \\
\hline
\hline\tstrut
\\
\begin{tabular}{c|c}
\hline\tstrut
Channel & $|g_i|$ \\
\hline\tstrut
& \\
$N\dvb$ & 3.4 \\
\hline
\end{tabular}  &
                      \begin{tabular}{c|c}
                      \hline\tstrut
                      Channel & $|g_i|$ \\
                      \hline\tstrut
                      & \\
                      $N\dsvb$ & 2.9 \\
                      $\Lambda\dvb$ & 2.0 \\
                      $\Sigma\dvb$ & 0.3 \\
                      $\sigs\db$ & 0.5 \\
                      $\sigs\dvb$ & 0.4 \\
                      \hline
                      \end{tabular} &
                                              \begin{tabular}{c|c}
                                              \hline\tstrut
                                              Channel & $|g_i|$ \\
                                              \hline\tstrut
                                              & \\
                                              $\Sigma\dsvb$ & 3.2 \\
                                              $\Xi\dvb$ & 3.0 \\
                                              $\sigs\dsb$ & $<$0.1 \\
                                              $\css\db$ & $<$0.1 \\
                                              $\sigs\dsvb$ & $<$0.1 \\
                                              $\css\dvb$ & $<$0.1 \\
                                              \hline
                                              \end{tabular} \\

\end{tabular}
\end{center}
\end{table}

\begin{table}
\begin{center}
\caption{Couplings of the poles belonging to the triplet of
Table \ref{tab:poles2}.} \label{tab:coup3}
\vspace{0.2cm}
\begin{tabular}{c||c}
\hline\tstrut
$S=-1$, $I=1/2$~~~~ & $S=-2$, $I=0$  \\
\hline
\hline\tstrut
\\
\begin{tabular}{c|c}
\hline\tstrut
Channel & $|g_i|$ \\
\hline\tstrut
& \\
$N\dsvb$ & 0.5 \\
$\Lambda\dvb$ & 1.8 \\
$\Sigma\dvb$ & 2.8 \\
$\sigs\db$ & 2.1 \\
$\sigs\dvb$ & 0.9 \\
\hline
\end{tabular}  &
                      \begin{tabular}{c|c}
                      \hline\tstrut
                      Channel & $|g_i|$ \\
                      \hline\tstrut
                      & \\
                      $\Lambda\dsvb$ & 2.5 \\
                      $\Xi\dvb$ & 3.2 \\
                      $\css\db$ & 1.1 \\
                      $\css\dvb$ & 0.7 \\
                      \hline
                      \end{tabular} \\

\end{tabular}
\end{center}
\end{table}
\begin{table}
\begin{center}
\caption{Couplings of the poles belonging to the heaviest 15-plet of
Table~\ref{tab:poles2}.} \label{tab:coup15he}
\vspace{0.2cm}
\begin{tabular}{ccc}
\hline\tstrut
$S=0$, $I=1$ & $S=-1$, $I=1/2$ & $S=-1$, $I=3/2$  \\
\hline
\hline\tstrut
\\
\begin{tabular}{c|c}
\hline\tstrut
Channel & $|g_i|$ \\
\hline\tstrut
& \\
$N\dvb$ & 1.3 \\
$\del\db$ & 1.2 \\
$\del\dvb$ & 4.4 \\
\hline
\end{tabular}  &
                      \begin{tabular}{c|c}
                      \hline \tstrut
                      Channel & $|g_i|$ \\
                      \hline\tstrut
                      & \\
                      $N\dsvb$ & 0.6 \\
                      $\Lambda\dvb$ & 0.8 \\
                      $\Sigma\dvb$ & 0.5 \\
                      $\sigs\db$ & 1.5 \\
                      $\sigs\dvb$ & 4.3\\
                      \hline 
                      \end{tabular} &
                                              \begin{tabular}{c|c}
                                              \hline\tstrut
                                              Channel & $|g_i|$ \\
                                              \hline\tstrut
                                              & \\
                                              $\del\dsb$ & 0.1 \\
                                              $\Sigma\dvb$ & 1.0 \\
                                              $\sigs\db$ & 1.7 \\
                                              $\del\dsvb$ & 3.9 \\
                                              $\sigs\dvb$ & 0.6 \\
                                               \hline    
                                              \end{tabular} \\

\\
\hline\tstrut
$S=-2$, $I=0$ & $S=-2$, $I=1$ & $S=-3$, $I=1/2$  \\
\hline
\hline\tstrut
\\
\begin{tabular}{c|c}
\hline\tstrut
Channel & $|g_i|$ \\
\hline\tstrut
& \\
$\Lambda\dsvb$ & 0.7 \\
$\Xi\dvb$ & 0.8 \\
$\css\db$ & 1.6 \\
$\css\dvb$ & 4.3 \\
\hline
\end{tabular}  &
                      \begin{tabular}{c|c}
                      \hline
                      Channel & $|g_i|$ \\
                      \hline\tstrut
                      & \\
                      $\Sigma\dsvb$ & 0.7 \\
                      $\Xi\dvb$ & 0.9 \\
                      $\sigs\dsb$ & 0.4 \\
                      $\css\db$ & 1.4 \\
                      $\sigs\dsvb$ & 4.0 \\
                      $\css\dvb$ & 1.0 \\
                      \hline
                      \end{tabular} &
                                              \begin{tabular}{c|c}
                                              \hline\tstrut
                                              Channel & $|g_i|$ \\
                                              \hline\tstrut
                                              & \\
                                              $\Xi\dsvb$ & 1.2 \\
                                              $\css\dsb$ & 0.4 \\
                                              $\Omega\db$ & 1.4 \\
                                              $\css\dsvb$ & 3.4 \\
                                              $\Omega\dvb$ & 2.6 \\
                                              \hline
                                              \end{tabular} \\

\end{tabular}
\end{center}
\end{table}

\subsection{The $J^P=\frac{5}{2}^-$ states}

The $J^P=\frac{5}{2}^-$ states found within our approach are shown in
Table~\ref{tab:poles3}.  There is only one possible combination of
multiplets with quantum numbers $J^P=\frac{5}{2}^-$ which is the
$10^*\otimes3^*$. From this interaction there is one attractive
15-plet. Most of the states in this sectors are bound by more than 100
MeV. Those are novel predictions from the present approach. We display in
Table \ref{tab:coup15f} the couplings of the generated resonances to
the different channels. 
\begin{table}
\begin{center}
\caption{Couplings of the poles belonging to the 15-plet of 
Table~\ref{tab:poles3}.} \label{tab:coup15f}
\vspace{0.2cm}
\begin{tabular}{ccc}
\hline\tstrut
$S=0$, $I=1$ ~~~~& $S=-1$, $I=1/2$~~~~ & $S=-1$, $I=3/2$  \\
\hline
\hline\tstrut
\\
\begin{tabular}{c|c}
\hline\tstrut
Channel & $|g_i|$ \\
\hline\tstrut
& \\
$\del\dvb$ & 4.2 \\
\hline
\end{tabular}  &
                      \begin{tabular}{c|c}
                      \hline\tstrut
                      Channel & $|g_i|$ \\
                      \hline\tstrut
                      & \\
                      $\sigs\dvb$ & 4.5 \\
                      \hline
                      \end{tabular} &
                                              \begin{tabular}{c|c}
                                              \hline\tstrut
                                              Channel & $|g_i|$ \\
                                              \hline\tstrut
                                              & \\
                                              $\del\dsvb$ & 3.8 \\
                                              $\sigs\dvb$ & 2.3 \\
                                              \hline
                                              \end{tabular} \\
\\
\hline\tstrut
$S=-2$, $I=0$~~~~ & $S=-2$, $I=1$ ~~~~~& $S=-3$, $I=1/2$  \\
\hline
\hline\tstrut
\\
\begin{tabular}{c|c}
\hline\tstrut
Channel & $|g_i|$ \\
\hline\tstrut
& \\
$\css\dvb$ & 4.8 \\
\hline
\end{tabular}  &
                      \begin{tabular}{c|c}
                      \hline\tstrut
                      Channel & $|g_i|$ \\
                      \hline\tstrut
                      & \\
                      $\sigs\dsvb$ & 3.3 \\
                      $\css\dvb$ & 3.3 \\
                      \hline
                      \end{tabular} &
                                              \begin{tabular}{c|c}
                                              \hline\tstrut
                                              Channel & $|g_i|$ \\
                                              \hline\tstrut
                                              & \\
                                              $\css\dsvb$ & 2.4 \\
                                              $\Omega\dvb$ & 4.1 \\
                                              \hline
                                              \end{tabular} \\

\end{tabular}
\end{center}
\end{table}


\section{Conclusions}\label{sec:concl}

In this work we have analyzed the spectrum of dynamically generated
resonances from the interaction of non-charmed baryons with negative
charmed mesons. In order to construct the interaction we followed a
model based on the $\rm{SU}(8)$ symmetry that respects the heavy quark
symmetry of the strong interactions in the charm sector. The
large spin--flavor symmetry is actually broken by the use of physical masses
for the hadrons and the decay constant of the $D$ mesons, $f_D$ and
$f_{D_s}$, instead of $f_\pi$ in the vertices.

Within the present approach we find a very rich spectrum of exotic
baryons with negative charm quantum number. Such states cannot be
constructed in the usual interpretation for the baryons whose
structure should be made of three quarks. With $J^P=\oh^-$ we find 6
$\rm{SU}(3)$ multiplets of resonances, namely a $15^\prime$-plet, two
15-plets, an anti-sextet and two triplets. For $J^P=\trh^-$ our model
generates 5 $\rm{SU}(3)$ multiplets, a $15^\prime$-plet, two 15-plets,
an anti-sextet and a triplet.  Finally for $J^P=\frac{5}{2}^-$ the
model generates a single 15-plet.

The results of the present approach were compared with previous
works. The $\rm{SU}(8)$ model generates the multiplets that had
already been predicted in \cite{hofmann,hofmann2,skyrme}, and the
states obtained here are around 100-200 MeV heavier than the exotic
states obtained in previous works. In addition, the  model
generates new multiplets on top of the ones predicted before. This is
a consequence of the inclusion of many more channels, as required to
enforce the heavy quark symmetry in the charm sector.

There has been one experimental claim for an anti-charmed baryon in
\cite{pentaq}. One of the dynamically generated states found within
the present approach could be a candidate for the anti-charmed baryon
experimentally claimed in \cite{pentaq}, although with a larger
width. In this case, we have also shown that there could be other decay channels
competing with the one analyzed in \cite{pentaq}.


\section*{Acknowledgments}

This work is partly supported by DGI and FEDER funds, under contract
FIS2006-03438, FIS2008-01143/FIS and PIE-CSIC 200850I238 and the Junta
de Andalucia grant no. FQM225. We acknowledge the support of the
European Community-Research Infrastructure Integrating Activity "Study
of Strongly Interacting Matter" (acronym HadronPhysics2, Grant
Agreement n. 227431) under the Seventh Framework Programme of EU.
Work supported in part by DFG (SFB/TR 16, "Subnuclear Structure of
Matter"). L.T. acknowledges support from the RFF program of the University of Groningen.


\newpage
\appendix
\section{The $\xi_{ij}^{CSIJ}$ matrices for $C=-1$}

\begin{table}[h]
\caption{
  $ C=-1$, $ S=-4$, $ I=0$, $ J= 1/2$.
}
\begin{ruledtabular}
\begin{tabular}{rrrrrrrrrrrrrrrrrrrrrrrrrrrrr}
 &  $ \Omega \DsSb $\\
$ \Omega \DsSb $& $ -2 $ \\
\end{tabular}
\end{ruledtabular}
\end{table}

\begin{table}[h]
\caption{
  $ C=-1$, $ S=-4$, $ I=0$, $ J= 3/2$.
}
\begin{ruledtabular}
\begin{tabular}{rrrrrrrrrrrrrrrrrrrrrrrrrrrrr}
 &  $ \Omega \Dsb $ &  $ \Omega 
  \DsSb $\\
$ \Omega \Dsb $& $ 3 $ 
 & $ \sqrt{ 15 } $ \\
$ \Omega \DsSb $
 & $ \sqrt{ 15 } $ 
 & $ 1 $ \\
\end{tabular}
\end{ruledtabular}
\end{table}
\begin{table}[h]
\caption{
  $ C=-1$, $ S=-4$, $ I=0$, $ J= 5/2$.
}
\begin{ruledtabular}
\begin{tabular}{rrrrrrrrrrrrrrrrrrrrrrrrrrrrr}
 &  $ \Omega \DsSb $\\
$ \Omega \DsSb $& $ 6 $ \\
\end{tabular}
\end{ruledtabular}
\end{table}
%
\begin{table}[h]
\caption{
  $ C=-1$, $ S=-3$, $ I=1/2$, $ J= 1/2$.
}
\begin{ruledtabular}
\begin{tabular}{rrrrrrrrrrrrrrrrrrrrrrrrrrrrr}
 &  $ \Cascada \Dsb $ 
 &  $ \Cascada \DsSb $ &  $ \CascadaS 
  \DsSb $ &  $ \Omega \DSb $\\
$ \Cascada \Dsb $& $ 2 $ 
 & $ \sqrt{\frac{ 16 }{ 3 }} 
   $ & $ \sqrt{\frac{ 8 }{ 3 
    }} $ & $ -\sqrt{ 8 } 
   $ \\
$ \Cascada \DsSb $
 & $ \sqrt{\frac{ 16 }{ 3 }} 
   $ & $ -\frac{ 2 }{ 3  } 
   $ & $ \sqrt{\frac{ 8 }{ 9 
    }} $ 
 & $ -\sqrt{\frac{ 8 }{ 3 }} 
   $ \\
$ \CascadaS \DsSb $
 & $ \sqrt{\frac{ 8 }{ 3 }} 
   $ & $ \sqrt{\frac{ 8 }{ 9 
    }} $ 
 & $ -\frac{ 4 }{ 3  } $ 
 & $ -\sqrt{\frac{ 4 }{ 3 }} 
   $ \\
$ \Omega \DSb $
 & $ -\sqrt{ 8 } $ 
 & $ -\sqrt{\frac{ 8 }{ 3 }} 
   $ & $ -\sqrt{\frac{ 4 }{ 3 
     }} $ & $ 0 $ \\
\end{tabular}
\end{ruledtabular}
\end{table}
\begin{table}[h]
\caption{
  $ C=-1$, $ S=-3$, $ I=1/2$, $ J= 3/2$.
}
\begin{ruledtabular}
\begin{tabular}{rrrrrrrrrrrrrrrrrrrrrrrrrrrrr}
 &  $ \Cascada \DsSb $ 
 &  $ \CascadaS \Dsb $ &  $ \Omega \Db 
  $ &  $ \CascadaS \DsSb $ 
 &  $ \Omega \DSb $\\
$ \Cascada \DsSb $
 & $ \frac{ 10 }{ 3  } $ 
 & $ -\sqrt{\frac{ 4 }{ 3 }} 
   $ & $ 2 $ 
 & $ \sqrt{\frac{ 20 }{ 9 }} 
   $ & $ -\sqrt{\frac{ 20 }{ 3 
     }} $ \\
$ \CascadaS \Dsb $
 & $ -\sqrt{\frac{ 4 }{ 3 }} 
   $ & $ 2 $ 
 & $ \sqrt{ 3 } $ 
 & $ \sqrt{\frac{ 20 }{ 3 }} 
   $ & $ \sqrt{ 5 } $ \\
$ \Omega \Db $& $ 2 $ 
 & $ \sqrt{ 3 } $ 
 & $ 0 $ & $ \sqrt{ 5 } 
   $ & $ 0 $ \\
$ \CascadaS \DsSb $
 & $ \sqrt{\frac{ 20 }{ 9 }} 
   $ & $ \sqrt{\frac{ 20 }{ 3 
    }} $ & $ \sqrt{ 5 } $ 
 & $ \frac{ 2 }{ 3  } $ 
 & $ \sqrt{\frac{ 1 }{ 3 }} 
   $ \\
$ \Omega \DSb $
 & $ -\sqrt{\frac{ 20 }{ 3 }} 
   $ & $ \sqrt{ 5 } $ 
 & $ 0 $ 
 & $ \sqrt{\frac{ 1 }{ 3 }} 
   $ & $ 0 $ \\
\end{tabular}
\end{ruledtabular}
\end{table}
\begin{table}[h]
\caption{
  $ C=-1$, $ S=-3$, $ I=1/2$, $ J= 5/2$.
}
\begin{ruledtabular}
\begin{tabular}{rrrrrrrrrrrrrrrrrrrrrrrrrrrrr}
 &  $ \CascadaS \DsSb $ 
 &  $ \Omega \DSb $\\
$ \CascadaS \DsSb $& $ 4 $ 
 & $ \sqrt{ 12 } $ \\
$ \Omega \DSb $
 & $ \sqrt{ 12 } $ 
 & $ 0 $ \\
\end{tabular}
\end{ruledtabular}
\end{table}
\begin{table}[h]
\caption{
  $ C=-1$, $ S=-2$, $ I=0$, $ J= 1/2$.
}
\begin{ruledtabular}
\begin{tabular}{rrrrrrrrrrrrrrrrrrrrrrrrrrrrr}
 &  $ \Lambda \Dsb $ &  $ \Cascada 
  \Db $ &  $ \Lambda \DsSb $ 
 &  $ \Cascada \DSb $ &  $ \CascadaS 
  \DSb $\\
$ \Lambda \Dsb $& $ 1 $ 
 & $ \sqrt{ 3 } $ 
 & $ \sqrt{ 3 } $ 
 & $ 1 $ & $ \sqrt{ 8 } 
   $ \\
$ \Cascada \Db $
 & $ \sqrt{ 3 } $ 
 & $ -1 $ & $ 1 $ 
 & $ \sqrt{\frac{ 1 }{ 3 }} 
   $ & $ \sqrt{\frac{ 8 }{ 3 
    }} $ \\
$ \Lambda \DsSb $
 & $ \sqrt{ 3 } $ 
 & $ 1 $ & $ -1 $ 
 & $ \sqrt{\frac{ 1 }{ 3 }} 
   $ & $ \sqrt{\frac{ 8 }{ 3 
    }} $ \\
$ \Cascada \DSb $& $ 1 $ 
 & $ \sqrt{\frac{ 1 }{ 3 }} 
   $ & $ \sqrt{\frac{ 1 }{ 3 
    }} $ 
 & $ -\frac{ 5 }{ 3  } $ 
 & $ \sqrt{\frac{ 8 }{ 9 }} 
   $ \\
$ \CascadaS \DSb $
 & $ \sqrt{ 8 } $ 
 & $ \sqrt{\frac{ 8 }{ 3 }} 
   $ & $ \sqrt{\frac{ 8 }{ 3 
    }} $ & $ \sqrt{\frac{ 8 
    }{ 9 }} $ 
 & $ \frac{ 2 }{ 3  } $ \\
\end{tabular}
\end{ruledtabular}
\end{table}
\begin{table}[h]
\caption{
  $ C=-1$, $ S=-2$, $ I=0$, $ J= 3/2$.
}
\begin{ruledtabular}
\begin{tabular}{rrrrrrrrrrrrrrrrrrrrrrrrrrrrr}
 &  $ \Lambda \DsSb $ 
 &  $ \Cascada \DSb $ &  $ \CascadaS 
  \Db $ &  $ \CascadaS \DSb $\\
$ \Lambda \DsSb $& $ 2 $ 
 & $ \sqrt{\frac{ 16 }{ 3 }} 
   $ & $ -2 $ 
 & $ \sqrt{\frac{ 20 }{ 3 }} 
   $ \\
$ \Cascada \DSb $
 & $ \sqrt{\frac{ 16 }{ 3 }} 
   $ & $ -\frac{ 2 }{ 3  } 
   $ & $ -\sqrt{\frac{ 4 }{ 3 
     }} $ & $ \sqrt{\frac{ 20 
    }{ 9 }} $ \\
$ \CascadaS \Db $& $ -2 $ 
 & $ -\sqrt{\frac{ 4 }{ 3 }} 
   $ & $ -1 $ 
 & $ -\sqrt{\frac{ 5 }{ 3 }} 
   $ \\
$ \CascadaS \DSb $
 & $ \sqrt{\frac{ 20 }{ 3 }} 
   $ & $ \sqrt{\frac{ 20 }{ 9 
    }} $ 
 & $ -\sqrt{\frac{ 5 }{ 3 }} 
   $ & $ -\frac{ 1 }{ 3  } 
   $ \\
\end{tabular}
\end{ruledtabular}
\end{table}
\begin{table}[h]
\caption{
  $ C=-1$, $ S=-2$, $ I=0$, $ J= 5/2$.
}
\begin{ruledtabular}
\begin{tabular}{rrrrrrrrrrrrrrrrrrrrrrrrrrrrr}
 &  $ \CascadaS \DSb $\\
$ \CascadaS \DSb $& $ -2 $ \\
\end{tabular}
\end{ruledtabular}
\end{table}
\begin{table}[h]
\caption{
  $ C=-1$, $ S=-2$, $ I=1$, $ J= 1/2$.
}
\begin{ruledtabular}
\begin{tabular}{rrrrrrrrrrrrrrrrrrrrrrrrrrrrr}
 &  $ \Sigma \Dsb $ &  $ \Cascada 
  \Db $ &  $ \Sigma \DsSb $ 
 &  $ \Cascada \DSb $ &  $ \SigmaS 
  \DsSb $ &  $ \CascadaS \DSb $\\
$ \Sigma \Dsb $& $ 1 $ 
 & $ 1 $ 
 & $ -\sqrt{\frac{ 1 }{ 3 }} 
   $ & $ \sqrt{\frac{ 25 }{ 3 
    }} $ & $ \sqrt{\frac{ 8 
    }{ 3 }} $ 
 & $ -\sqrt{\frac{ 8 }{ 3 }} 
   $ \\
$ \Cascada \Db $& $ 1 $ 
 & $ 1 $ 
 & $ \sqrt{\frac{ 25 }{ 3 }} 
   $ & $ -\sqrt{\frac{ 1 }{ 3 
     }} $ & $ \sqrt{\frac{ 8 
    }{ 3 }} $ 
 & $ -\sqrt{\frac{ 8 }{ 3 }} 
   $ \\
$ \Sigma \DsSb $
 & $ -\sqrt{\frac{ 1 }{ 3 }} 
   $ & $ \sqrt{\frac{ 25 }{ 3 
    }} $ & $ \frac{ 5 }{ 3 
     } $ 
 & $ -\frac{ 7 }{ 3  } $ 
 & $ \sqrt{\frac{ 8 }{ 9 }} 
   $ & $ -\sqrt{\frac{ 8 }{ 9 
     }} $ \\
$ \Cascada \DSb $
 & $ \sqrt{\frac{ 25 }{ 3 }} 
   $ & $ -\sqrt{\frac{ 1 }{ 3 
     }} $ 
 & $ -\frac{ 7 }{ 3  } $ 
 & $ \frac{ 5 }{ 3  } $ 
 & $ \sqrt{\frac{ 8 }{ 9 }} 
   $ & $ -\sqrt{\frac{ 8 }{ 9 
     }} $ \\
$ \SigmaS \DsSb $
 & $ \sqrt{\frac{ 8 }{ 3 }} 
   $ & $ \sqrt{\frac{ 8 }{ 3 
    }} $ & $ \sqrt{\frac{ 8 
    }{ 9 }} $ 
 & $ \sqrt{\frac{ 8 }{ 9 }} 
   $ & $ -\frac{ 2 }{ 3  } 
   $ & $ -\frac{ 4 }{ 3  } 
   $ \\
$ \CascadaS \DSb $
 & $ -\sqrt{\frac{ 8 }{ 3 }} 
   $ & $ -\sqrt{\frac{ 8 }{ 3 
     }} $ 
 & $ -\sqrt{\frac{ 8 }{ 9 }} 
   $ & $ -\sqrt{\frac{ 8 }{ 9 
     }} $ 
 & $ -\frac{ 4 }{ 3  } $ 
 & $ -\frac{ 2 }{ 3  } $ \\
\end{tabular}
\end{ruledtabular}
\end{table}
\begin{table}[h]
\caption{
  $ C=-1$, $ S=-2$, $ I=1$, $ J= 3/2$.
}
\begin{ruledtabular}
\begin{tabular}{rrrrrrrrrrrrrrrrrrrrrrrrrrrrr}
 &  $ \Sigma \DsSb $ &  $ \Cascada 
  \DSb $ &  $ \SigmaS \Dsb $ 
 &  $ \CascadaS \Db $ &  $ \SigmaS 
  \DsSb $ &  $ \CascadaS \DSb $\\
$ \Sigma \DsSb $
 & $ \frac{ 2 }{ 3  } $ 
 & $ \frac{ 8 }{ 3  } $ 
 & $ -\sqrt{\frac{ 4 }{ 3 }} 
   $ & $ \sqrt{\frac{ 4 }{ 3 
    }} $ & $ \sqrt{\frac{ 20 
    }{ 9 }} $ 
 & $ -\sqrt{\frac{ 20 }{ 9 }} 
   $ \\
$ \Cascada \DSb $
 & $ \frac{ 8 }{ 3  } $ 
 & $ \frac{ 2 }{ 3  } $ 
 & $ -\sqrt{\frac{ 4 }{ 3 }} 
   $ & $ \sqrt{\frac{ 4 }{ 3 
    }} $ & $ \sqrt{\frac{ 20 
    }{ 9 }} $ 
 & $ -\sqrt{\frac{ 20 }{ 9 }} 
   $ \\
$ \SigmaS \Dsb $
 & $ -\sqrt{\frac{ 4 }{ 3 }} 
   $ & $ -\sqrt{\frac{ 4 }{ 3 
     }} $ & $ 1 $ 
 & $ 2 $ 
 & $ \sqrt{\frac{ 5 }{ 3 }} 
   $ & $ \sqrt{\frac{ 20 }{ 3 
    }} $ \\
$ \CascadaS \Db $
 & $ \sqrt{\frac{ 4 }{ 3 }} 
   $ & $ \sqrt{\frac{ 4 }{ 3 
    }} $ & $ 2 $ & $ 1 $ 
 & $ \sqrt{\frac{ 20 }{ 3 }} 
   $ & $ \sqrt{\frac{ 5 }{ 3 
    }} $ \\
$ \SigmaS \DsSb $
 & $ \sqrt{\frac{ 20 }{ 9 }} 
   $ & $ \sqrt{\frac{ 20 }{ 9 
    }} $ & $ \sqrt{\frac{ 5 
    }{ 3 }} $ 
 & $ \sqrt{\frac{ 20 }{ 3 }} 
   $ & $ \frac{ 1 }{ 3  } 
   $ & $ \frac{ 2 }{ 3  } 
   $ \\
$ \CascadaS \DSb $
 & $ -\sqrt{\frac{ 20 }{ 9 }} 
   $ & $ -\sqrt{\frac{ 20 }{ 9 
     }} $ & $ \sqrt{\frac{ 20 
    }{ 3 }} $ 
 & $ \sqrt{\frac{ 5 }{ 3 }} 
   $ & $ \frac{ 2 }{ 3  } 
   $ & $ \frac{ 1 }{ 3  } 
   $ \\
\end{tabular}
\end{ruledtabular}
\end{table}
\begin{table}[h]
\caption{
  $ C=-1$, $ S=-2$, $ I=1$, $ J= 5/2$.
}
\begin{ruledtabular}
\begin{tabular}{rrrrrrrrrrrrrrrrrrrrrrrrrrrrr}
 &  $ \SigmaS \DsSb $ 
 &  $ \CascadaS \DSb $\\
$ \SigmaS \DsSb $& $ 2 $ 
 & $ 4 $ \\
$ \CascadaS \DSb $& $ 4 $ 
 & $ 2 $ \\
\end{tabular}
\end{ruledtabular}
\end{table}
\begin{table}[h]
\caption{
  $ C=-1$, $ S=-1$, $ I=1/2$, $ J= 1/2$.
}
\begin{ruledtabular}
\begin{tabular}{rrrrrrrrrrrrrrrrrrrrrrrrrrrrr}
 &  $ \Nucleon \Dsb $ 
 &  $ \Lambda \Db $ &  $ \Nucleon \DsSb 
  $ &  $ \Sigma \Db $ 
 &  $ \Lambda \DSb $ &  $ \Sigma \DSb 
  $ &  $ \SigmaS \DSb $\\
$ \Nucleon \Dsb $& $ 0 $ 
 & $ \sqrt{\frac{ 3 }{ 2 }} 
   $ & $ 0 $ 
 & $ \sqrt{\frac{ 3 }{ 2 }} 
   $ & $ \sqrt{\frac{ 9 }{ 2 
    }} $ 
 & $ -\sqrt{\frac{ 1 }{ 2 }} 
   $ & $ 2 $ \\
$ \Lambda \Db $
 & $ \sqrt{\frac{ 3 }{ 2 }} 
   $ & $ 1 $ 
 & $ \sqrt{\frac{ 9 }{ 2 }} 
   $ & $ 0 $ & $ 0 $ 
 & $ \sqrt{ 3 } $ 
 & $ \sqrt{ 6 } $ \\
$ \Nucleon \DsSb $& $ 0 $ 
 & $ \sqrt{\frac{ 9 }{ 2 }} 
   $ & $ 0 $ 
 & $ -\sqrt{\frac{ 1 }{ 2 }} 
   $ & $ -\sqrt{\frac{ 3 }{ 2 
     }} $ & $ \sqrt{\frac{ 25 
    }{ 6 }} $ 
 & $ \sqrt{\frac{ 4 }{ 3 }} 
   $ \\
$ \Sigma \Db $
 & $ \sqrt{\frac{ 3 }{ 2 }} 
   $ & $ 0 $ 
 & $ -\sqrt{\frac{ 1 }{ 2 }} 
   $ & $ -1 $ 
 & $ \sqrt{ 3 } $ 
 & $ -\sqrt{\frac{ 4 }{ 3 }} 
   $ & $ \sqrt{\frac{ 2 }{ 3 
    }} $ \\
$ \Lambda \DSb $
 & $ \sqrt{\frac{ 9 }{ 2 }} 
   $ & $ 0 $ 
 & $ -\sqrt{\frac{ 3 }{ 2 }} 
   $ & $ \sqrt{ 3 } $ 
 & $ 1 $ & $ -2 $ 
 & $ \sqrt{ 2 } $ \\
$ \Sigma \DSb $
 & $ -\sqrt{\frac{ 1 }{ 2 }} 
   $ & $ \sqrt{ 3 } $ 
 & $ \sqrt{\frac{ 25 }{ 6 }} 
   $ & $ -\sqrt{\frac{ 4 }{ 3 
     }} $ & $ -2 $ 
 & $ \frac{ 1 }{ 3  } $ 
 & $ \sqrt{\frac{ 2 }{ 9 }} 
   $ \\
$ \SigmaS \DSb $& $ 2 $ 
 & $ \sqrt{ 6 } $ 
 & $ \sqrt{\frac{ 4 }{ 3 }} 
   $ & $ \sqrt{\frac{ 2 }{ 3 
    }} $ & $ \sqrt{ 2 } $ 
 & $ \sqrt{\frac{ 2 }{ 9 }} 
   $ & $ \frac{ 2 }{ 3  } 
   $ \\
\end{tabular}
\end{ruledtabular}
\end{table}
\begin{table}[h]
\caption{
  $ C=-1$, $ S=-1$, $ I=1/2$, $ J= 3/2$.
}
\begin{ruledtabular}
\begin{tabular}{rrrrrrrrrrrrrrrrrrrrrrrrrrrrr}
 &  $ \Nucleon \DsSb $ 
 &  $ \Lambda \DSb $ &  $ \Sigma \DSb 
  $ &  $ \SigmaS \Db $ 
 &  $ \SigmaS \DSb $\\
$ \Nucleon \DsSb $& $ 0 $ 
 & $ \sqrt{ 6 } $ 
 & $ \sqrt{\frac{ 2 }{ 3 }} 
   $ & $ -\sqrt{ 2 } $ 
 & $ \sqrt{\frac{ 10 }{ 3 }} 
   $ \\
$ \Lambda \DSb $
 & $ \sqrt{ 6 } $ 
 & $ 1 $ & $ 1 $ 
 & $ -\sqrt{ 3 } $ 
 & $ \sqrt{ 5 } $ \\
$ \Sigma \DSb $
 & $ \sqrt{\frac{ 2 }{ 3 }} 
   $ & $ 1 $ 
 & $ -\frac{ 5 }{ 3  } $ 
 & $ -\sqrt{\frac{ 1 }{ 3 }} 
   $ & $ \sqrt{\frac{ 5 }{ 9 
    }} $ \\
$ \SigmaS \Db $
 & $ -\sqrt{ 2 } $ 
 & $ -\sqrt{ 3 } $ 
 & $ -\sqrt{\frac{ 1 }{ 3 }} 
   $ & $ -1 $ 
 & $ -\sqrt{\frac{ 5 }{ 3 }} 
   $ \\
$ \SigmaS \DSb $
 & $ \sqrt{\frac{ 10 }{ 3 }} 
   $ & $ \sqrt{ 5 } $ 
 & $ \sqrt{\frac{ 5 }{ 9 }} 
   $ & $ -\sqrt{\frac{ 5 }{ 3 
     }} $ 
 & $ -\frac{ 1 }{ 3  } $ \\
\end{tabular}
\end{ruledtabular}
\end{table}
\begin{table}[h]
\caption{
  $ C=-1$, $ S=-1$, $ I=1/2$, $ J= 5/2$.
}
\begin{ruledtabular}
\begin{tabular}{rrrrrrrrrrrrrrrrrrrrrrrrrrrrr}
 &  $ \SigmaS \DSb $\\
$ \SigmaS \DSb $& $ -2 $ \\
\end{tabular}
\end{ruledtabular}
\end{table}
\begin{table}[h]
\caption{
  $ C=-1$, $ S=-1$, $ I=3/2$, $ J= 1/2$.
}
\begin{ruledtabular}
\begin{tabular}{rrrrrrrrrrrrrrrrrrrrrrrrrrrrr}
 &  $ \Sigma \Db $ &  $ \Sigma \DSb 
  $ &  $ \Delta \DsSb $ 
 &  $ \SigmaS \DSb $\\
$ \Sigma \Db $& $ 2 $ 
 & $ \sqrt{\frac{ 16 }{ 3 }} 
   $ & $ \sqrt{ 8 } $ 
 & $ -\sqrt{\frac{ 8 }{ 3 }} 
   $ \\
$ \Sigma \DSb $
 & $ \sqrt{\frac{ 16 }{ 3 }} 
   $ & $ -\frac{ 2 }{ 3  } 
   $ & $ \sqrt{\frac{ 8 }{ 3 
    }} $ 
 & $ -\sqrt{\frac{ 8 }{ 9 }} 
   $ \\
$ \Delta \DsSb $
 & $ \sqrt{ 8 } $ 
 & $ \sqrt{\frac{ 8 }{ 3 }} 
   $ & $ 0 $ 
 & $ -\sqrt{\frac{ 4 }{ 3 }} 
   $ \\
$ \SigmaS \DSb $
 & $ -\sqrt{\frac{ 8 }{ 3 }} 
   $ & $ -\sqrt{\frac{ 8 }{ 9 
     }} $ 
 & $ -\sqrt{\frac{ 4 }{ 3 }} 
   $ & $ -\frac{ 4 }{ 3  } 
   $ \\
\end{tabular}
\end{ruledtabular}
\end{table}
\begin{table}[h]
\caption{
  $ C=-1$, $ S=-1$, $ I=3/2$, $ J= 3/2$.
}
\begin{ruledtabular}
\begin{tabular}{rrrrrrrrrrrrrrrrrrrrrrrrrrrrr}
 &  $ \Delta \Dsb $ &  $ \Sigma \DSb 
  $ &  $ \SigmaS \Db $ 
 &  $ \Delta \DsSb $ &  $ \SigmaS \DSb 
  $\\
$ \Delta \Dsb $& $ 0 $ 
 & $ -2 $ 
 & $ \sqrt{ 3 } $ 
 & $ 0 $ & $ \sqrt{ 5 } 
   $ \\
$ \Sigma \DSb $& $ -2 $ 
 & $ \frac{ 10 }{ 3  } $ 
 & $ \sqrt{\frac{ 4 }{ 3 }} 
   $ & $ \sqrt{\frac{ 20 }{ 3 
    }} $ 
 & $ -\sqrt{\frac{ 20 }{ 9 }} 
   $ \\
$ \SigmaS \Db $
 & $ \sqrt{ 3 } $ 
 & $ \sqrt{\frac{ 4 }{ 3 }} 
   $ & $ 2 $ 
 & $ \sqrt{ 5 } $ 
 & $ \sqrt{\frac{ 20 }{ 3 }} 
   $ \\
$ \Delta \DsSb $& $ 0 $ 
 & $ \sqrt{\frac{ 20 }{ 3 }} 
   $ & $ \sqrt{ 5 } $ 
 & $ 0 $ 
 & $ \sqrt{\frac{ 1 }{ 3 }} 
   $ \\
$ \SigmaS \DSb $
 & $ \sqrt{ 5 } $ 
 & $ -\sqrt{\frac{ 20 }{ 9 }} 
   $ & $ \sqrt{\frac{ 20 }{ 3 
    }} $ & $ \sqrt{\frac{ 1 
    }{ 3 }} $ 
 & $ \frac{ 2 }{ 3  } $ \\
\end{tabular}
\end{ruledtabular}
\end{table}
\begin{table}[h]
\caption{
  $ C=-1$, $ S=-1$, $ I=3/2$, $ J= 5/2$.
}
\begin{ruledtabular}
\begin{tabular}{rrrrrrrrrrrrrrrrrrrrrrrrrrrrr}
 &  $ \Delta \DsSb $ &  $ \SigmaS 
  \DSb $\\
$ \Delta \DsSb $& $ 0 $ 
 & $ \sqrt{ 12 } $ \\
$ \SigmaS \DSb $
 & $ \sqrt{ 12 } $ 
 & $ 4 $ \\
\end{tabular}
\end{ruledtabular}
\end{table}
\begin{table}[h]
\caption{
  $ C=-1$, $ S=0$, $ I=0$, $ J= 1/2$.
}\label{tab:tab1mev}
\begin{ruledtabular}
\begin{tabular}{rrrrrrrrrrrrrrrrrrrrrrrrrrrrr}
 &  $ \Nucleon \Db $ &  $ \Nucleon 
  \DSb $\\
$ \Nucleon \Db $& $ 0 $ 
 & $ -\sqrt{ 12 } $ \\
$ \Nucleon \DSb $
 & $ -\sqrt{ 12 } $ 
 & $ 4 $ \\
\end{tabular}
\end{ruledtabular}
\end{table}
\begin{table}[h]
\caption{
  $ C=-1$, $ S=0$, $ I=0$, $ J= 3/2$.
}
\begin{ruledtabular}
\begin{tabular}{rrrrrrrrrrrrrrrrrrrrrrrrrrrrr}
 &  $ \Nucleon \DSb $\\
$ \Nucleon \DSb $& $ -2 $ \\
\end{tabular}
\end{ruledtabular}
\end{table}
\begin{table}[h]
\caption{
  $ C=-1$, $ S=0$, $ I=1$, $ J= 1/2$.
}
\begin{ruledtabular}
\begin{tabular}{rrrrrrrrrrrrrrrrrrrrrrrrrrrrr}
 &  $ \Nucleon \Db $ &  $ \Nucleon 
  \DSb $ &  $ \Delta \DSb $\\
$ \Nucleon \Db $& $ 2 $ 
 & $ \sqrt{\frac{ 16 }{ 3 }} 
   $ & $ \sqrt{\frac{ 32 }{ 3 
    }} $ \\
$ \Nucleon \DSb $
 & $ \sqrt{\frac{ 16 }{ 3 }} 
   $ & $ -\frac{ 2 }{ 3  } 
   $ & $ \sqrt{\frac{ 32 }{ 9 
    }} $ \\
$ \Delta \DSb $
 & $ \sqrt{\frac{ 32 }{ 3 }} 
   $ & $ \sqrt{\frac{ 32 }{ 9 
    }} $ & $ \frac{ 2 }{ 3 
     } $ \\
\end{tabular}
\end{ruledtabular}
\end{table}
\begin{table}[h]
\caption{
  $ C=-1$, $ S=0$, $ I=1$, $ J= 3/2$.
}
\begin{ruledtabular}
\begin{tabular}{rrrrrrrrrrrrrrrrrrrrrrrrrrrrr}
 &  $ \Nucleon \DSb $ 
 &  $ \Delta \Db $ &  $ \Delta \DSb 
  $\\
$ \Nucleon \DSb $
 & $ \frac{ 10 }{ 3  } $ 
 & $ -\sqrt{\frac{ 16 }{ 3 }} 
   $ & $ \sqrt{\frac{ 80 }{ 9 
    }} $ \\
$ \Delta \Db $
 & $ -\sqrt{\frac{ 16 }{ 3 }} 
   $ & $ -1 $ 
 & $ -\sqrt{\frac{ 5 }{ 3 }} 
   $ \\
$ \Delta \DSb $
 & $ \sqrt{\frac{ 80 }{ 9 }} 
   $ & $ -\sqrt{\frac{ 5 }{ 3 
     }} $ 
 & $ -\frac{ 1 }{ 3  } $ \\
\end{tabular}
\end{ruledtabular}
\end{table}
\begin{table}[h]
\caption{
  $ C=-1$, $ S=0$, $ I=1$, $ J= 5/2$.
}
\begin{ruledtabular}
\begin{tabular}{rrrrrrrrrrrrrrrrrrrrrrrrrrrrr}
 &  $ \Delta \DSb $\\
$ \Delta \DSb $& $ -2 $ \\
\end{tabular}
\end{ruledtabular}
\end{table}
\begin{table}[h]
\caption{
  $ C=-1$, $ S=0$, $ I=2$, $ J= 1/2$.
}
\begin{ruledtabular}
\begin{tabular}{rrrrrrrrrrrrrrrrrrrrrrrrrrrrr}
 &  $ \Delta \DSb $\\
$ \Delta \DSb $& $ -2 $ \\
\end{tabular}
\end{ruledtabular}
\end{table}
\begin{table}[h]
\caption{
  $ C=-1$, $ S=0$, $ I=2$, $ J= 3/2$.
}
\begin{ruledtabular}
\begin{tabular}{rrrrrrrrrrrrrrrrrrrrrrrrrrrrr}
 &  $ \Delta \Db $ &  $ \Delta \DSb 
  $\\
$ \Delta \Db $& $ 3 $ 
 & $ \sqrt{ 15 } $ \\
$ \Delta \DSb $
 & $ \sqrt{ 15 } $ 
 & $ 1 $ \\
\end{tabular}
\end{ruledtabular}
\end{table}

\begin{table}[h]
\caption{
  $ C=-1$, $ S=0$, $ I=2$, $ J= 5/2$.
}
\begin{ruledtabular}
\begin{tabular}{rrrrrrrrr}
 &  $ \Delta \DSb $\\
$ \Delta \DSb $& $ 6 $ \\
\end{tabular}
\end{ruledtabular}
\end{table}


\begin{thebibliography}{99}

\bibitem{exp1}
  M.~Artuso {\it et al.}  [CLEO Collaboration],
  Phys.\ Rev.\ Lett.\  {\bf 86}, 4479 (2001)
  [arXiv:hep-ex/0010080].

\bibitem{exp2}
  R.~Mizuk {\it et al.}  [Belle Collaboration],
  Phys.\ Rev.\ Lett.\  {\bf 94}, 122002 (2005)
  [arXiv:hep-ex/0412069].

\bibitem{exp3}
  C.~P.~Jessop {\it et al.}  [CLEO Collaboration],
  Phys.\ Rev.\ Lett.\  {\bf 82}, 492 (1999)
  [arXiv:hep-ex/9810036].

\bibitem{exp4}
  S.~E.~Csorna {\it et al.}  [CLEO Collaboration],
  Phys.\ Rev.\ Lett.\  {\bf 86}, 4243 (2001)
  [arXiv:hep-ex/0012020].

\bibitem{exp5}
  R.~Chistov {\it et al.}  [BELLE Collaboration],
  Phys.\ Rev.\ Lett.\  {\bf 97}, 162001 (2006)
  [arXiv:hep-ex/0606051].

\bibitem{hid1}
  T.~Amirzadeh {\it et al.}  [Birmingham-CERN-Glasgow-Michigan State-Paris
                  Collaboration and Cambridge-Mi],
  Phys.\ Lett.\  B {\bf 89}, 125 (1979).

\bibitem{hid2}
  V.~M.~Karnaukhov, V.~I.~Moroz, C.~Coca and A.~Mihul,
  Phys.\ Lett.\  B {\bf 281}, 148 (1992).

\bibitem{pentaq}
  A.~Aktas {\it et al.}  [H1 Collaboration],
  Phys.\ Lett.\  B {\bf 588}, 17 (2004)
  [arXiv:hep-ex/0403017].
  
\bibitem{lutz}
  M.~F.~M.~Lutz and E.~E.~Kolomeitsev,
  Nucl.\ Phys.\  A {\bf 730}, 110 (2004)
  [arXiv:hep-ph/0307233].

\bibitem{estela}
  C.~E.~Jimenez-Tejero, A.~Ramos and I.~Vidana,
  Phys.\ Rev.\  C {\bf 80}, 055206 (2009)
  [arXiv:0907.5316 [hep-ph]].
  
\bibitem{hofmann}
  J.~Hofmann and M.~F.~M.~Lutz,
  Nucl.\ Phys.\  A {\bf 763}, 90 (2005)
  [arXiv:hep-ph/0507071].
  
\bibitem{hofmann2}
  J.~Hofmann and M.~F.~M.~Lutz,
  Nucl.\ Phys.\  A {\bf 776}, 17 (2006)
  [arXiv:hep-ph/0601249].

\bibitem{haidenbauer}
  J.~Haidenbauer, G.~Krein, U.~G.~Meissner and A.~Sibirtsev,
  Eur.\ Phys.\ J.\  A {\bf 33} (2007) 107
  [arXiv:0704.3668 [nucl-th]].

\bibitem{juan1}
  C.~Garcia-Recio, V.~K.~Magas, T.~Mizutani, J.~Nieves, A.~Ramos, L.~L.~Salcedo and L.~Tolos,
  Phys.\ Rev.\  D {\bf 79}, 054004 (2009)
  [arXiv:0807.2969 [hep-ph]].

\bibitem{recio}
  C.~Garcia-Recio, J.~Nieves and L.~L.~Salcedo,
  Phys.\ Rev.\  D {\bf 74}, 036004 (2006)
  [arXiv:hep-ph/0605059].

\bibitem{juan2}
  C.~Garcia-Recio, J.~Nieves and L.~L.~Salcedo,
  Phys.\ Rev.\  D {\bf 74}, 034025 (2006)
  [arXiv:hep-ph/0505233].

\bibitem{Toki:2007ab}
  H.~Toki, C.~Garcia-Recio and J.~Nieves,
  Phys.\ Rev.\  D {\bf 77} (2008) 034001
  [arXiv:0711.3536 [hep-ph]].
 


\bibitem{juan3}
  J.~Nieves and E.~Ruiz Arriola,
  Phys.\ Rev.\  D {\bf 64}, 116008 (2001)
  [arXiv:hep-ph/0104307].
  
\bibitem{rocaev}
  L.~Roca, E.~Oset and J.~Singh,
  Phys.\ Rev.\  D {\bf 72}, 014002 (2005)
  [arXiv:hep-ph/0503273].

\bibitem{meusca}
  D.~Gamermann, E.~Oset, D.~Strottman and M.~J.~Vicente Vacas,
  Phys.\ Rev.\  D {\bf 76}, 074016 (2007)
  [arXiv:hep-ph/0612179].

\bibitem{meuwave}
  D.~Gamermann, J.~Nieves, E.~Oset and E.~R.~Arriola,
   Phys.\ Rev.\  D {\bf 81} (2010) 014029
  arXiv:1001.3254 [hep-ph].

  
\bibitem{skyrme}
  B.~Wu and B.~Q.~Ma,
  Phys.\ Rev.\  D {\bf 70}, 034025 (2004)
  [arXiv:hep-ph/0402244].
  
\bibitem{roca}
  L.~Roca, S.~Sarkar, V.~K.~Magas and E.~Oset,
  Phys.\ Rev.\  C {\bf 73}, 045208 (2006)
  [arXiv:hep-ph/0603222].

\bibitem{other}
  B.~Aubert {\it et al.}  [BABAR Collaboration],
  Phys.\ Rev.\  D {\bf 73}, 091101 (2006)
  [arXiv:hep-ex/0604006].
  
\bibitem{meuax}
  D.~Gamermann and E.~Oset,
  Eur.\ Phys.\ J.\  A {\bf 33}, 119 (2007)
  [arXiv:0704.2314 [hep-ph]].



\end{thebibliography}
\end{document}